\documentclass[twocolumn,showpacs,preprintnumbers,amsmath,amssymb,superscriptaddress]{revtex4}
\usepackage{bm}% bold math  twocolumn,
\usepackage{epsfig,amssymb,amsmath,bm}
\begin{document}

%\preprint{APS/123-QED}

\title{Hypercomplex Dirac Equation and Electrodynamics of Non-Conserved Charges.}
\author{K. S. Karplyuk}
 \affiliation{Department of Radiophysics, Taras Shevchenko University, Academic 
Glushkov prospect 2, building 5, Kyiv 03122, Ukraine, Phone: +(380 44)-264-7171}

\author{O. O. Zhmudskyy}%
\email{ozhmudsky@physics.ucf.edu}
\affiliation{%
Department of Physics, University of Central Florida,  4000 Central Florida Blvd. 
Orlando, FL, 32816 Phone: (407)-823-6941
}% 

\begin{abstract}
It is shown that the hypercomplex Dirac equation describes the
system of connected fields: 4-scalar, 4-pseudoscalar, 4-vector,
4-pseudo\-vec\-tor and antisymmetric 4-tensor second rank field.
If mass is assumed to be  zero this system splits into  
two subsystems. Equations containing tensor, scalar and
pseudoscalar fields coincide with Maxwell equations complemented
by scalar and pseudoscalar fields. This system describes the
electrodynamics of non-conserved charges. The scalar and
pseudoscalar fields are generated only by the non-conserved
charges --- electric and hypothetical magnetic. The influence  
of these fields on the charged particles is very unusual --- it
causes a change of their rest mass. This allows us to give a new
look at the Wigner paradox and mechanism of mass
renormalization.
\end{abstract}


\maketitle
%2345678901234567890123456789012345678901234567890123456789012345678901234567890
\section{Introduction}
It is known that Maxwell equations can be written down as the
massless Dirac equation if the elements of dirac column are
represented as $\psi_i=a_jE_j+b_kB_k$. Here $a_j$ and $b_k$ are
some coefficients and $E_j$ and $B_k$ are components of the
electric and magnetic fields. Such representation can be done in several ways so
different authors have different combination for $a_jE_j+b_kB_k$.
The first versions of such a notation were proposed in
\cite{dar,lap,op} and later this subject was developed in
several hundreds of works.

There is another method to use the Dirac equation for the
derivation of new equations. According to this method the 
$\psi$-function is considered as a hyper\-comp\-lex number built on
16 independent dirac matrices $\varGamma^i$. We will study this approach below 
and we will show that presentation of the 
$\psi$-function as $\psi=a_i\varGamma^i$ results in the system of
equations which describes the set of mutually connected fields of
different tensor dimensional representation.

The obtained system of equations contains some known equations.
Among them, the Maxwell equations complemented by the scalar
field $\epsilon$. Such modified Maxwell equations are used for
electromagnetic field quantization~\cite{heis}. The introduction  
of a scalar field however always was considered as a purely formal
technique necessary for implementation of the canonical
quantization procedure. The physi\-cal nature of the scalar field
has not been studied. It turns out however that extended in such a
way, the Maxwell equations describe the electrodynamics of the
non-conserved charges. A scalar $\epsilon$-field is generated only by the 
non-conserved charges. It can spread out forming a wave together
with the longitudinal electric field. Its influence on the charged
particles is very unusual --- it changes the rest mass of these
particles. In particular, self-action of this field on the
non-conserved charges, generating them, results in transformation of
their rest energy (i.e. rest mass) into the energy of their
electric field and vice versa. It becomes possible to observe in what way part 
of the mass of an initially uncharged particle transforms into electromagnetic
mass while a charge is generated for this particle.  
 Thus we have an opportunity to compare the physical
mechanism of re\-normali\-za\-tion with the idea of such mass
renormalization. It is possible as well to avoid contradictions
during consideration of processes with the non-conserved charges
mentioned by Wigner. Below we will consider these and other
questions in more detail.

%2345678901234567890123456789012345678901234567890123456789012345678901234567890
\section{Hypercomplex Dirac equation}
It is known that the system of hypercomplex numbers with 16 units
can be built on dirac matrices as follows: 

\[\gamma^\mu\gamma^\nu+\gamma^\nu\gamma^\mu=2\eta^{\mu\nu}.\]

Here, $\eta^{\mu\nu}$ is the matrix of Minkowski metrical tensor,
$\eta^{\mu\nu} =\rm{diag}(1,-1,-1,-1)$. The formal solution of the
Dirac equation

\begin{equation}
\left(i\gamma^\nu\partial_\nu-\frac{mc}{\hbar}\right)\psi=0
\label{Eq:01}
\end{equation}

may be treated as a hypercomplex number of this system. From a 
group point of view such a solution will belong to the reduced
presentation of the Lorentz group. The importance of such
solutions was pointed out by Lev \cite{lev}. Let us discuss this
possibility.

We will look for a solution of equation (\ref{Eq:01}) of the form

\begin{eqnarray}
\psi=\epsilon I+iV_\nu\gamma^\nu+F_{0k}\gamma^0\gamma^k+
F_{23}\gamma^2\gamma^3+ F_{31}\gamma^3\gamma^1 \nonumber \\
+F_{12}\gamma^1\gamma^2 +iU_0\gamma^1\gamma^2\gamma^3 + 
iU_1\gamma^0\gamma^2\gamma^3 \nonumber \\
+iU_2\gamma^0\gamma^3\gamma^1+ iU_3\gamma^0\gamma^1\gamma^2+
\beta\gamma^0\gamma^1\gamma^2\gamma^3.
\label{Eq:02}
\end{eqnarray}

%2345678901234567890123456789012345678901234567890123456789012345678901234567890
Here, $I$ is the unity matrix, the Latin indices take on values
1,2,3, Greek indices - 0,1,2,3.  
Repeated Latin indices are summed from 0 to 3.  
Repeated Greek indices are summed from 0 to 4.  Substituting $\psi$ into 
(\ref{Eq:01}) and equating the coefficients of each of the 16 independent 
matrices to zero we get 16 equations:

\begin{eqnarray}
\frac{\partial\epsilon}{\partial x^\nu}-\frac{\partial
F_\nu^{{}\,\,\mu}}{\partial x^\mu}-\varkappa V_\nu&=0,  \label{Eq:03}  \\
\frac{\partial\beta}{\partial x^\nu}+\frac{\eta_{\nu\alpha}}{2}
\varepsilon^{\alpha\beta\gamma\delta}\frac{\partial
F_{\gamma\delta}}{\partial x^\beta}-\varkappa U_\nu&=0, \label{Eq:04} \\
\frac{\partial V_\alpha}{\partial x^\beta}- \frac{\partial
V_\beta}{\partial x^\alpha}+\eta_{\alpha\nu}\eta_{\beta\mu}
\varepsilon^{\nu\mu\gamma\delta}\frac{\partial U_\gamma}{\partial
x^\delta}-\varkappa F_{\alpha\beta}&=0, \label{Eq:05} \\
 \frac{\partial V^\nu}{\partial x^\nu}+\varkappa\epsilon&=0,  \label{Eq:06} \\
\frac{\partial U^\nu}{\partial x^\nu}+\varkappa\beta&=0.  \label{Eq:07} \\
\nonumber
\end{eqnarray}

Here, $\varkappa$ is the Compton wave number,
$\varkappa=2\pi/\lambda_k=mc/\hbar$.

Equations (\ref{Eq:03})-(\ref{Eq:07})  describe the system of connected 
fields: 4-scalar $\epsilon$, 4-pseudoscalar $\beta$, 4-vector $V_\nu$,
4-pseudovector $U_\nu$ and an antisymmetric 4-tensor second rank
field $F_{\mu\nu}$. In the case of zero-point mass ($\varkappa=0$)
this system splits into two independent subsystems.
Equations (\ref{Eq:03}),(\ref{Eq:04}) in this case coincide with 
homogeneous Maxwell equations complemented by scalar filed $\epsilon$ and 
pseudoscalar field $\beta$. Equations (\ref{Eq:05})-(\ref{Eq:07}) describe the 
system of connected 4-vector and 4-pseudovector fields.
%2345678901234567890123456789012345678901234567890123456789012345678901234567890

Equations (\ref{Eq:03})-(\ref{Eq:07}) are homogeneous. We put the sources of
the fields into right sides of the Eq. (\ref{Eq:03})-(\ref{Eq:07}) in order to study 
connections between the fields and their sources as well as the influence of the
fields on corresponding charges and currents.  In three-dimensional notation a
suitable system looks like

\begin{eqnarray}
\frac{1}{c}\frac{\partial {\epsilon}}{\partial t}+\nabla\cdot{\bm E} - 
\varkappa V^0&=\zeta c\rho_e, \label{Eq:08} \\ 
\frac{1}{c}\frac{\partial {\bm E}}{\partial t}-\nabla\times c{\bm B} + 
\nabla\epsilon +\varkappa{\bm V}&=-\zeta{\bm j}_e, \label{Eq:09} \\ 
-\frac{1}{c}\frac{\partial {\beta}}{\partial t}+\nabla\cdot c{\bm B} + 
\varkappa U^0&=-\zeta c\rho_m, \label{Eq:10} \\ 
\frac{1}{c}\frac{\partial c{\bm B}}{\partial t}+\nabla\times {\bm E} - 
\nabla\beta - \varkappa{\bm U}&=\zeta{\bm j}_m, \label{Eq:11} \\
\frac{1}{c}\frac{\partial V^0}{\partial t}+\nabla\cdot{\bm V} + 
\varkappa{\epsilon}&=\zeta s, \label{Eq:12} \\ 
\frac{1}{c}\frac{\partial {\bm V}}{\partial t}-\nabla\times {\bm U} + 
\nabla V^0-\varkappa{\bm E}&=-\zeta{\bm k}, \label{Eq:13} \\ 
\frac{1}{c}\frac{\partial {\bm U}}{\partial t}+\nabla\times {\bm V} + 
\nabla U^0+ \varkappa c{\bm B}&=\zeta{\bm l}, \label{Eq:14} \\
\frac{1}{c}\frac{\partial  U^0}{\partial t}+\nabla\cdot{\bm U} + 
\varkappa{\beta}&=-\zeta p.  \label{Eq:15} \\
\nonumber 
\end{eqnarray}

Identification of three-dimensional and four-dimensional coordinates
of vectors and tensors are as usuall:
$\bm{E}=(F_{01},F_{02},F_{03})$, $c\bm{B}=(-F_{23},$
$-F_{31},-F_{12})$, $\bm{V}=(V^1,V^2,V^3)$,
$\bm{U}=(U^1,U^2,U^3)$. Equations (\ref{Eq:08})-(\ref{Eq:15}) are written down in SI
system but instead of constants $\varepsilon_0$ and $\mu_0$ we
used velocity of light $c=1/\sqrt{\varepsilon_0\mu_0}$ and
resistance of the vacuum $\zeta=\sqrt{\mu_0/\varepsilon_0}$. This form
of notation as applied to Maxwell equations allows us to use all the 
advantages of the Gaussian unit system, remaining at the same time in SI unit
system.

In the system (\ref{Eq:08})-(\ref{Eq:15}): $\rho_e$ and ${\bm j}_e$  are the
densities of electric charges and currents, $\rho_m$ and ${\bm
j}_m$ are the densities of hypothetical magnetic charges and
currents. The nature of the other sources is to be determined. By
analogy with an electric current
$j^{\mu}_e=e\bar{\psi}\gamma^{\mu}\psi$ we can introduce 
bilinear combinations for the other sources, $s=e\bar{\psi}\psi$,
$k^i=e\bar{\psi}\gamma^0\gamma^i\psi$,
$l^i=e\bar{\psi}\gamma^j\gamma^k\psi$,
$p=e\bar{\psi}\gamma^5\psi$,
$j^{\mu}_m=e\bar{\psi}\gamma^{\mu}\gamma^5\psi$. The investigation
of such a supposition helps us to find out why these bilinear
combinations are not the sources of the fields and do not generate
interactions like the $j^{\mu}=e\bar{\psi}\gamma^{\mu}\psi$.

The fields can be expressed using potentials:

\begin{eqnarray}
\epsilon&=&\frac{1}{c}\frac{\partial \varphi}{\partial t} + 
\nabla\cdot c{\bm A}+\varkappa S, \label{Eq:16} \\ 
{\bm E}&=& -\frac{1}{c}\frac{\partial c{\bm A}}{\partial t} - 
\nabla \varphi+\nabla\times\bm{\varPi}+\varkappa\bm{\theta}, \label{Eq:17} \\ 
c{\bm B}&=&\frac{1}{c}\frac{\partial\bm{\varPi}}{\partial t} + 
\nabla\varPi^0+\nabla\times c{\bm A}+ \varkappa\bm{\vartheta}, \label{Eq:18} \\ 
\beta&=&\frac{1}{c}\frac{\partial \varPi^0}{\partial t} + 
\nabla\cdot\bm{\varPi}- \varkappa\varPsi, \label{Eq:19} \\
V^0&=&\frac{1}{c}\frac{\partial S}{\partial t} + 
\nabla\cdot\bm{\theta}-\varkappa \varphi, \label{Eq:20} \\ 
{\bm V }&=&-\frac{1}{c}\frac{\partial\bm{\theta}}{\partial t} + 
\nabla\times\bm{\vartheta}- \nabla S-\varkappa c{\bm A}, \label{Eq:21} \\
U^0&=&-\frac{1}{c}\frac{\partial{\varPsi}}{\partial t} - 
\nabla\cdot\bm{\vartheta}-\varkappa\varPi^0, \label{Eq:22} \\ 
{\bm U }&=&\frac{1}{c}\frac{\partial\bm{\vartheta}}{\partial t} + 
\nabla\times\bm{\theta}+ \nabla\varPsi-\varkappa\bm{\varPi}. \label{Eq:23}  \\
\nonumber
\end{eqnarray}

%2345678901234567890123456789012345678901234567890123456789012345678901234567890
Substitution of these expressions into (\ref{Eq:08})-(\ref{Eq:15}), yields to 
the following equations for the potentials:

\begin{eqnarray} 
\square \varphi+\varkappa^2 \varphi&=&\zeta c\rho_e,\quad 
\square c\bm{A}+\varkappa^2 c\bm{A}=\zeta\bm{j}_e, \label{Eq:24} \\ 
\square \varPi^0+\varkappa^2 \varPi^0&=&\zeta c\rho_m,\quad\:  
\square \bm{\varPi}+\varkappa^2 \bm{\varPi}=\zeta\bm{j}_m, \label{Eq:25} \\ 
\square \bm{\theta}+\varkappa^2 \bm{\theta}&=&\zeta\bm{k},\quad\qquad   
\square \bm{\vartheta}+\varkappa^2 \bm{\vartheta}=\zeta\bm{l}, \label{Eq:26} \\ 
\square S+\varkappa^2S&=&\zeta s,\quad\qquad   
\square\varPsi+\varkappa^2\varPsi=\zeta p. \label{Eq:27} \\
\nonumber
\end{eqnarray}

Here, $\square$ is the d'Alambertian,
$\square=c^{-2}\partial^2/\partial t^2-\triangle$.

As for the Maxwell electro\-dy\-na\-mics it is possible to get conservation 
laws of energy and momentum from equations (\ref{Eq:08})-(\ref{Eq:15}). 
The derivation of these conservation laws is presented in the appendix A.

Equations (\ref{Eq:08})-(\ref{Eq:15}) include some well known systems of 
equations as particular cases.   We get the Proka equations for the 
homogeneous system (\ref{Eq:24})-(\ref{Eq:27}) if we also set up 
$\epsilon=\beta=\bm{U}=U^0=0$.  In the case $\varkappa=0$, system 
(\ref{Eq:08})-(\ref{Eq:15}) splits into two independent 
systems. One of these subsystems coincides
with the Maxwell equations comple\-ment\-ed by fields $\epsilon$ and
$\beta$.

%2345678901234567890123456789012345678901234567890123456789012345678901234567890
The main question to be answered about equations (\ref{Eq:08})-(\ref{Eq:15})
is whether these equations have any reference to physical
reality? In order to answer this we have to investigate the
properties of these equations.

\section{Electrodynamics of the non-conserved charges}
It is known that the Maxwell equations describe only such
electromagnetic processes in which electric charge is conserved. 
On this subject Feynman wrote \cite{feyn}: {\em \lq\lq The
laws of physics have no answer to the question: \lq\lq What
happens if a charge is suddenly created at this point --- what
electromagnetic effects are produced\rq\rq? No answer can be given
because our equations say it doesn't happen. If it were to happen,
we would need new laws, but we cannot say what they would
be.\rq\rq}

It turns out that equations (\ref{Eq:08})-(\ref{Eq:15}) also describe, 
among others, processes  with non-conserved electric charges. If 
$\varkappa=0$ these equations split into two independent systems.
The first one (equations (\ref{Eq:08})-(\ref{Eq:11})) are the Maxwell
equations complemented by the fields $\epsilon$ and $\beta$. 

Let us eliminate for a while hypothetical magnetic
charges. In order to do this we must set  
$\rho_m=0$, $\bm{j}_m=0$, $\beta=0$.  In this case we have:

\begin{eqnarray} 
\frac{1}{c}\frac{\partial {\epsilon}}{\partial t} + 
\nabla\cdot\bm{E}&=&\zeta\rho c, \label{Eq:28} \\
-\frac{1}{c}\frac{\partial\bm{E}}{\partial t} + 
\nabla\times{c}\bm{B}-\nabla\epsilon&=&\zeta \bm{j}, \label{Eq:29} \\ 
\nabla\cdot{c}\bm{B}&=&0, \label{Eq:30} \\
\frac{1}{c}\frac{\partial{c}\bm{B}}{\partial t} + 
\nabla\times\bm{E}&=& 0. \label{Eq:31}  \\
\nonumber
\end{eqnarray}

As was noted by Heisenberg and Pauli \cite{heis}, in order to quantize the 
electromagnetic field we can introduce the $\epsilon$-field into Maxwell 
equations.  It is known that the application of canonical quantization procedure 
to the Maxwell equations has not succeeded, because a zero canonical momentum
corresponds to the canonical coordinate $\varphi$. Introduction of the 
$\epsilon$-field which is just a
canonical momentum for a coordinate $\varphi$ solves this problem. After
implementation of quantization to equations (\ref{Eq:28})-(\ref{Eq:31})  
the mean value of the $\epsilon$-field is set to zero,   
$\langle\psi|\epsilon|\psi\rangle=0$. Thus the $\epsilon$-field is
removed from consideration and we go back to the Maxwell equations. As
the $\epsilon$-field plays only a formal role in this procedure its
physical nature was not studied. 

%2345678901234567890123456789012345678901234567890123456789012345678901234567890
\subsection{Sources of $\epsilon$-field}
 First of all, we need to find the source of the $\epsilon$-field. Let us 
take the derivative $c^{-1}\partial/\partial t$ of equation (\ref{Eq:28}) and  
the divergence of equation (\ref{Eq:29}), and add them:

\begin{equation}
\square\epsilon=\zeta\left(\frac{\partial \rho}{\partial t} + 
\nabla\cdot\bm{j}\right).
\label{Eq:32}
\end{equation}

The right hand side of the equation (\ref{Eq:32}) is a source of the $\epsilon$-field. 
It differs from zero only if the conservation law of
electric charge is not valid,

\[\frac{\partial\rho}{\partial t} + \nabla\cdot\bm{j}\neq 0.\] 

Thus, the $\epsilon$-field is generated only by the non-conserved 
electric charges.

The $\epsilon$-field can form a wave propagating jointly with
the longitudinal electric field. The plane $\epsilon-\bm{E}$ wave has the
form

\begin{eqnarray} 
\epsilon=\epsilon_0\cos(\omega t-\bm{k}\cdot\bm{r}),\qquad
\bm{E}=\bm{k}\frac{c}{\omega}\epsilon_0\cos(\omega t-\bm{k}\cdot\bm{r}), \nonumber \\
\bm{B}=0, \qquad  
k^2=\frac{\omega^2}{c^2}.\nonumber \\
\nonumber 
\end{eqnarray} 

Unlike a transversal wave this wave does not contain the magnetic field. 
The electric field of such a wave is directed along the phase velocity,
$\bm{E}\parallel\bm{k}$. We will illustrate the process of
$\epsilon$-field birth by an example. Let us find the
$\epsilon$-field generated by a spherical shell of radius $r_0$.  
Let us also assume that charge is distributed uniformly over the sphere 
and its time dependence $q(t)$ obeys an exponential law.  
We chose as an example a sphere with a finite radius, instead of point charge, 
to avoid infinity at $r=0$. The potential $\varphi$ of such a shell in
the external region $r\geq r_0$ is determined by the retarded
solution of equation (\ref{Eq:24}):

\[\varphi(r,t)=\frac{\zeta
c}{4\pi}\frac{1}{2rr_0}\int_{r-r_0}^{r+r_0}q\left(t-\frac{x}{c}\right)dx.\]

When the charge increases from 0 to $q_0$ according to:
%2345678901234567890123456789012345678901234567890123456789012345678901234567890
 
\begin{equation} 
q(t)=\left \{
\begin{array}{cl}
0  &t\leq 0,\\ q_0(1-e^{-\displaystyle{t\over\tau}})\hspace{7mm} &t\geq 0,
\end{array} 
\right.
\label{Eq:33}
\end{equation}

the potential looks like

\begin{equation} 
\varphi_{\scriptscriptstyle\uparrow} = \frac{\zeta c}{4\pi}\frac{q_0}{2rr_0}\times   \nonumber
\end{equation} 

\begin{equation} 
\left \{ 
\begin{array}{cl} 0 \quad & t\leq \displaystyle{r-r_0\over c},
\\{ct-c\tau+r_0-r} + {c\tau}e^{r-r_0-ct\over c\tau} \quad & 
\displaystyle{r-r_0\over c}\leq t\leq \displaystyle{r+r_0\over c}, 
\\ 2r_0-{c\tau} \left(e^{r+r_0\over c\tau} - 
e^{r-r_0\over c\tau}\right) e^{-{t\over \tau}}\quad &
\displaystyle{r+r_0\over c}\leq t<\infty.
\end{array}
\right .  \nonumber 
\end{equation}

The corresponding $\epsilon$-field for this potential is: 

\begin{eqnarray}
\epsilon_{\scriptscriptstyle\uparrow}(r,t) = \frac{1}{c}
\frac{\partial\varphi_{\scriptscriptstyle\uparrow}}{\partial t} = \frac{\zeta 
c}{4\pi}\frac{q_0}{2rr_0}\times \nonumber \\ 
\left \{
\begin{array}{cl}
0 \hspace{2mm} &t\leq \displaystyle\frac{r-r_0}{c},\\
1-e^{\frac{r-r_0}{c\tau}}e^{-\frac{t}{\tau}}
 \hspace{2mm} & \displaystyle\frac{r-r_0}{c}\leq t\leq \frac{r+r_0}{c},\\
\left( e^{\frac{r+r_0}{c\tau}}-
e^{\frac{r-r_0}{c\tau}}\right)e^{-\frac{t}{\tau}}\hspace{2mm}
&\displaystyle\frac{r+r_0}{c}\leq t<\infty,
\end{array}
\right .   
\label{Eq:34}
\end{eqnarray} 

The corresponding longitudinal electric field for this potential is: 
%2345678901234567890123456789012345678901234567890123456789012345678901234567890
\begin{equation}
E_{r{\scriptscriptstyle\uparrow}}(r,t)=-\frac{\partial\varphi_{
\scriptscriptstyle\uparrow}}{\partial r}=\epsilon_{\scriptscriptstyle\uparrow}
(r,t)+\frac{\varphi_{\scriptscriptstyle\uparrow}}{r}.
\label{Eq:35}
\end{equation}

We can see that the $\epsilon$-field (\ref{Eq:34}) has the form of a 
spherical layer --- it is concentrated in a spherical layer with an
external radius of $r=r_0+ct$ and width $c\tau$. This spherical layer 
propagates at the speed of light. As the spherical layer moves away 
from the charged shell the $\epsilon$-field decreases as $1/r$. The
electric field consists of two parts. One part has the same form
as the $\epsilon$-field. This is a leading front of the electric
field propagating away from the shell. The coulomb field which is 
formed behind this front is the second part of the electric field. 

When the charge of the shell decreases from $q_0$ according to  
\begin{equation}
q(t)=\left \{
\begin{array}{cl}
q_0  &t\leq 0,\\ q_0e^{-\displaystyle{t\over \tau}}\hspace{5mm} &t\geq 0,
\end{array}
\right .
\label{Eq:36}
\end{equation}

its potential in the external region $r\geq r_0$ is given by 

\begin{equation}
\varphi_{\scriptscriptstyle\downarrow}(r,t) = \frac{\zeta c}{4\pi}
\frac{q_0}{2rr_0}\times 
\nonumber
\end{equation}
\begin{equation}
\left \{\begin{array}{cl} 2r_0 \hspace{4mm} &t\leq \displaystyle\frac{r-r_0}{c},\\
{c\tau-ct+r_0+r} - {c\tau}e^{\frac{r-r_0-ct}{c\tau}} \hspace{4mm} &
\displaystyle\frac{r-r_0}{c}\leq t\leq \displaystyle\frac{r+r_0}{c},\\ {c\tau}
\left(e^{\frac{r+r_0}{c\tau}}-e^{\frac{r-r_0}{c\tau}}\right)e^{-\frac{t}{\tau}}
\hspace{4mm}&\displaystyle\frac{r+r_0}{c}\leq t<\infty.
\end{array}
\right .  \nonumber
\end{equation}

%2345678901234567890123456789012345678901234567890123456789012345678901234567890
Such a vanishing charge creates the $\epsilon$-field
\begin{equation}
\epsilon_{\scriptscriptstyle\downarrow}(r,t)=
-\epsilon_{\scriptscriptstyle\uparrow}(r,t)
\label{Eq:37}
\end{equation}

and the longitudinal electric field

\begin{equation}
E_{r{\scriptscriptstyle\downarrow}}(r,t)=-\frac{\partial
\varphi_{\scriptscriptstyle\downarrow}}{\partial
r}=\epsilon_{\scriptscriptstyle\downarrow}(r,t)+
\frac{\varphi_{\scriptscriptstyle\downarrow}}{r}.
\label{Eq:38}
\end{equation}

The $\epsilon$-field created by a vanishing charge is the same
spherical layer as the field of the charge being generated, but
has an opposite sign. The electric field consists of two parts.
One part is the same spherical layer as the $\epsilon$-field. This
layer propagates away from a vanishing charge and \lq\lq erases\rq\rq\
its coulomb field. Behind the spherical layer the coulomb field has 
been erased, but it is still extant in front of the spherical layer. 
This is the second part of the electric field.

\subsection{Influence of the $\epsilon$-field on charges}
What is the effect of $\epsilon$-field on other charges? In order to answer 
this question let us turn the conservation laws of energy and
momentum (\ref{Eq:56}),(\ref{Eq:57}). For equations (\ref{Eq:28})-(\ref{Eq:31}) 
they are:

\begin{equation}
\frac{\partial}{\partial t}
\left(\frac{E^2+c^2B^2+\epsilon^2}{2\zeta
c}\right)+(\bm{E}\bm{j}-\epsilon\rho c)
=-\mathrm{div}\frac{\bm{E}\times{\sf
c}\bm{B}+\epsilon\bm{E}}{\zeta}.
\label{Eq:39}
\end{equation}

\begin{equation}\frac{\partial}{\partial t}\frac
{\bm{E}\times c\bm{B}-\epsilon\bm{E}}{\zeta c^2}+
\left(\rho\bm{E}+\frac{1}{c}\bm{j}\times c\bm{B}-
\frac{1}{c}\epsilon\bm{j}\right)=\nabla\cdot\mathsf{T}.
\label{Eq:40}
\end{equation}

%2345678901234567890123456789012345678901234567890123456789012345678901234567890
The terms $\rho\bm{E}+\bm{j}\times \bm{B}- \epsilon\bm{j}/c$ in
(\ref{Eq:40}) describe forces which are exerted on the charges, and
the terms $\bm{E}\bm{j}-\epsilon\rho c$ in (\ref{Eq:39}) describe  the power of
these forces. For a point charge $q$ these
terms describe the rate of change of energy
$\mathcal{E}=mc^2/\sqrt{1-v^2/c^2}$ and change of momentum
$\bm{p}=m\bm{v}/\sqrt{1-v^2/c^2}$ of this charge:

\begin{equation}
\frac{d\mathcal{E}}{dt}=q\bm{v}\cdot\bm{E}-qc\epsilon=
q\bm{v}\cdot\bm{E}-
\epsilon\mathcal{E}\frac{q}{mc}\sqrt{1-\frac{v^2}{c^2}},
\label{Eq:41}
\end{equation}

\begin{equation}
\frac{d\bm{p}}{dt}=q\bm{E}+\frac{q}{c}\bm{v}\times
c\bm{B}-\frac{q}{c}\epsilon\bm{v}=q\bm{E}+\frac{q}{c}\bm{v}\times
c\bm{B}-\epsilon\bm{p}\frac{q}{mc}\sqrt{1-\frac{v^2}{c^2}}.
\label{Eq:42}
\end{equation}

As we can see $\epsilon$-field brings a contribution to the 
interaction. The contri\-bu\-tion is significantly  different from the 
contributions of the electric field $\bm{E}$ and magnetic field $\bm{B}$. 
Fields $\bm{E}$ and $\bm{B}$ rotate the 4-vector of energy-momentum of the charged
particle in Minkowski space without changing its magnitude. Conversely, 
the $\epsilon$-field, as we can see from (\ref{Eq:41})-(\ref{Eq:42}), changes
all coordinates of energy-momentum vector proportionally, i.e.   
changes magnitude of this vector. In other words, the
$\epsilon$-field changes the mass of the affected charge. Let us derive an 
equation describing this change.   In order to do this let us take product of 
Eq. (\ref{Eq:41}) with $2\mathcal{E}$, and the dot product of Eq. (\ref{Eq:42}) 
with $2\bm{p}c^2$, and subtract them:  

\begin{equation}
\frac{d(\mathcal{E}^2-p^2c^2)}{dt}=\frac{d\,m^2c^4}{dt}= \nonumber
\end{equation}
\begin{equation}
=2q\left(\mathcal{E}\bm{v}\cdot\bm{E}-c^2\bm{p}\cdot\bm{E}-
c\bm{p}\cdot\bm{v}\times с \bm{B}-c\mathcal{E}\epsilon+
c\bm{p}\cdot\bm{v}\epsilon\right) = \nonumber 
\end{equation}
\begin{equation}
-2qc^3\epsilon m\sqrt{1-\frac{v^2}{c^2}}. \nonumber
\end{equation}

This yields:

\begin{equation}
\frac{dmc^2}{dt}=
-qc\epsilon\sqrt{1-\frac{v^2}{c^2}},\hspace{9mm}\mbox{or}\hspace{9mm}
\frac{dmc^2}{d\tau}=-qc\epsilon.  \label{Eq:43}
\end{equation}

This is the rate of mass change of the charged particle by the interaction 
term $\epsilon j^{\mu}$.

For a system of charges distributed with density  $\rho$, which is moving 
with velocity $v$, Eq. (\ref{Eq:43}) must be changed as follows:

%If the $\epsilon$-field affect not a point charge, but affect the
%charges distributed with density $\rho$ which charges move as a
%whole with a speed $v$, then a change of the mass of this
%distribution is described by equation
\begin{equation}
\frac{dmc^2}{dt}= -\int_Vc\rho\epsilon\sqrt{1-\frac{v^2}{c^2}}dV.
\label{Eq:44} 
\end{equation}

Thus we find that a varying charge creates the $\epsilon$-field, 
which affects other charges and changes their rest masses. When 
a charge which is being created acts on a charge of the same sign, 
by means of the $\epsilon$-field, the rest mass of the second charge  
decreases. The rate of this decreasing is proportional to the 
rate of charge increasing and value of the charge exposed to the 
$\epsilon$-field.

At a glance such action of the $\epsilon$-field seems unusual.
However, upon close consideration it turns out to be natural and
resonable. Let us illustrate this by the example of a constant 
charge $q$, which interacts with a varying  rest charge
$q_1$. Charge energy change is expressed by  (\ref{Eq:41}):

\[\frac{d\mathcal{E}}{dt}=q\bm{v}\cdot\bm{E}-qc\epsilon=
-q\frac{\partial
\varphi_1}{\partial\bm{r}}\frac{d\bm{r}}{dt}-q\frac{\partial
\varphi_1}{\partial t}=-q\frac{d\varphi_1(\bm{r})}{dt},\] 

or

\begin{equation}
\frac{m(t)c^2}{\sqrt{1-{v^2/c^2}}}+ q\varphi_1(t,\bm{r})=const.
\label{Eq:45} 
\end{equation}

One can see that any change in the potential energy of charge $q$  
generated by interaction with $q_1$ is compensated by its energy
change  $mc^2/\sqrt{1-v^2/c^2}$. 
It becomes trivial if charge $q_1$ does not change and creates only the electric field. 
For example, charge $q$ at infinity moving towards $q_1$, with 
velocity $v$, approaches $q_1$ until its
primary kinetic energy $mc^2/\sqrt{1-v^2/c^2}-mc^2$ becomes equal
to potential energy $q\varphi_1$. However, for varying charge $q_1$, the 
potential energy of charge $q$ can change for reasons other than its motion 
relative to $q_1$. For example, increasing charge $q_1$
creates the fields $\epsilon$ and $\bm{E}$. While
propagating they achieve  charge $q$ which appears in the
scope of influence of potential $\varphi_1$. Therefore its potential
energy changes from zero to $q\varphi_1$. The rest and
constant charge $q$ has only one possibility to compensate this
change, in agreement with the conservation law of energy, --- to change 
its rest energy and mass. This 
takes place under the influence  of the $\epsilon$-field and is described by
the terms  $\epsilon\rho c$ and $\epsilon\bm{j}/c$ in
equations (\ref{Eq:41}) and (\ref{Eq:42}).

\subsection{Radiation of energy by non-conserved charge}
We find that varying charge is at rest generate $\epsilon$
and $\bm{E}$ fields. These fields propagate away from the charge and
carry some energy. According to (\ref{Eq:39}) the flux density of this
energy is equal to $\epsilon\bm{E}/\zeta$. As an example we will find
the energy emitted by a spherical shell with a charge
changing according to Eq. (\ref{Eq:33}) or Eq. (\ref{Eq:36}). We have already found
the fields of such a charged spherical shell of radius $r_0$. Let us surround this shell by a
spherical surface of radius $R$ $(R \geq r_0)$ and calculate the
energy flux through this surface for  the time interval 
from 0 to $\infty$:

\begin{eqnarray}
\int_0^\infty\frac{\epsilon E}{\zeta}4\pi
R^2dt&=&\pm\frac{c\zeta}{4\pi}\frac{q_0^2}{2R} \nonumber \\
&+&\frac{c\zeta}{4\pi}\frac{q_0^2}{2r_0}
\left[1-\frac{c\tau}{2r_0}\left(1-e^{-\frac{2r_0}{c\tau}}\right)\right]. \nonumber \\ 
\label{Eq:46} 
\end{eqnarray}

The sign \lq\lq+\rq\rq\ refers to the case of charge increasing (\ref{Eq:33}),
sign \lq\lq--\rq\rq\ refers to the case of charge
decreasing (\ref{Eq:36}).

The first term in Eq.~(\ref{Eq:46}) is equal in magnitude to the coulomb
field energy in the range  $R\leq r<\infty$ outside the surface $R$.
The coulomb energy of the increasing charge propagates away from it and 
flows through the surface $R$  into the volume $r\geq R$ (the first
component with the sign \lq\lq+\rq\rq\ ).  The coulomb energy from 
$r\geq R$ flows into the volume $r\leq R$ through the surface $R$ 
and goes back to the decreasing charge (the first component with 
the sign \lq\lq--\rq\rq ).   

If $R\to \infty$ the first term tends to zero, i.e., the coulomb
energy does not pass through an infinitely distant surface. It is
\lq\lq tied\rq\rq\ to the charge and it is not emitted
irretrievably.

%2345678901234567890123456789012345678901234567890123456789012345678901234567890
The second term in Eq.~(\ref{Eq:46}) does not depend on surface radius $R$. 
It describes energy which is carried off irretrievably by
the leading front impulses of the electric and
$\epsilon$-fields. When these impulses pass through the 
surface of radius $R$ surrounding the charge, they carry energy through it  
described by the second component in (\ref{Eq:46}). If the radius of the
charged shell is small enough, such that $r_0\ll c\tau$, the second
component tends to the value

\begin{equation}
W_{rad}=\frac{\zeta c}{4\pi}\frac{q_0^2}{2c\tau}.
\label{Eq:47} 
\end{equation}

Such energy is emitted by the point charge when it increases or
decreases according to the Eq. (\ref{Eq:33}) or Eq. (\ref{Eq:36}).

The natural question is: what is the source of the coulomb energy, which 
propagates away from the charge and disappears forever? In order to answer 
this question, let us investigate the energy balance while the charge is varying.
For this purpose we use the integral form of Eq. (\ref{Eq:39}):

\begin{equation}
\frac{d}{dt}\int_V \frac{E^2+\epsilon^2}{2\zeta c}{dV}-\int_V\epsilon c
\rho(\bm{r})dV = -\int_S\frac{\epsilon\bm{E}}{\zeta}d\bm{S}.
\label{Eq:48} 
\end{equation}

This equation expresses energy balance of stationary charge 
distribution during the change of $\rho$ in volume $V$, enclosed by 
the surface of radius $R$.  Using (\ref{Eq:44}), we find

\begin{equation}
-\frac{d\,mc^2}{dt}=\frac{d}{dt}\int_V\frac{E^2+\epsilon^2}{2\zeta c}{dV} + 
\int_S\frac{\epsilon\bm{E}}{\zeta}d\bm{S}.
\label{Eq:49} 
\end{equation}

It is easy to see that the energy source of electric and $\epsilon$-fields 
is the rest energy of the particles with varying charges.

Let us apply Eq. (\ref{Eq:49}) to the above example of
the charged shell follow what happens in this case. Let us calculate the energy change due to 
Eq. (\ref{Eq:48}) inside the spherical surface of radius $R$,
enclosing the $r_0$ shell, during all the time of shell charge
varying, from 0 to $\infty$.

Charge increases according to Eq. (\ref{Eq:33}) create an 
$\epsilon$-field (\ref{Eq:34}). This $\epsilon$-field affects shell charges
and changes their rest energies according to (\ref{Eq:44})

\[-\Delta mc^2=\int_0^{\infty}\left(\int_Vc\rho\epsilon_{
\scriptscriptstyle\uparrow} dV\right)dt = c\int_0^{\infty}q(t)\epsilon_{
\scriptscriptstyle\uparrow}(r_0,t)dt=\]

\[=\frac{\zeta c}{4\pi}
\frac{q_0^2}{2r_0}+\frac{\zeta c}{4\pi}\frac{q_0^2}{2r_0}
\left[1-\frac{c\tau}{2r_0}\left(1-e^{-\frac{2r_o}{c\tau}}\right)\right].  
\eqno (49a) 
\]

While the shell charge increases to $q_0$ its primary rest energy $m_0c^2$
decreases by the amount given in (49a).   Let us calculate varying energy of the fields.   
There are no fields inside the $R$ surface at $t=0$. 
There is the coulomb field between this surface and the $r_0$ surface when $t\to\infty$,
which has an energy equal to 
$\frac{\zeta c}{4\pi}\left(\frac{q_0^2}{2r_0}-\frac{q_0^2}{2R}\right)$. 
Moreover, the energy given in (\ref{Eq:46}) has been emitted through the surface $R$. Thus,
equation (\ref{Eq:48}) integrated over time from 0 to $\infty$ takes on the form

\[\underbrace{\frac{\zeta c}{4\pi} \frac{q_0^2}{2r_0}+\frac{\zeta
c}{4\pi}\frac{q_0^2}{2r_0} \left[1-\frac{c\tau}{2r_0}
\left(1-e^{-\frac{2r_o}{c\tau}}\right)\right]}_{{{-\Delta
mc^2}}}=\]

%2345678901234567890123456789012345678901234567890123456789012345678901234567890
\begin{eqnarray}
=\underbrace{\frac{\zeta c}{4\pi}
\left(\frac{q_0^2}{2r_0}-\frac{q_0^2}{2R}\right)}_{{\scriptsize{W_{Coul}
}}} + \nonumber \\ 
\underbrace{\frac{\zeta c}{4\pi}\frac{q_0^2}{2r_0}
\left[1-\frac{c\tau}{2r_0}
\left(1-e^{-\frac{2r_o}{c\tau}}\right)\right]+\frac{\zeta c}{4\pi}
\frac{q_0^2}{2R}}_{{\scriptsize{W_{rad} }}}.
\label{Eq:50} 
\end{eqnarray}

One can see that at charge birth its rest energy decreases,
and due to this it acquires  
the energy of the coulomb field.  I.e., the energy of the coulomb field, propagating 
away from a charge during its birth, is taken from its rest 
energy. Thus, the birth of a charge is the effective method of direct 
transformation of rest energy into electric energy. Moreover, at charge birth 
 a finite part of rest energy is emitted irretrievably. The
decrease of rest energy is exactly equal to the sum of the coulomb field
energy and emitted energy.

In the case when $r_0\ll c\tau$, $R\to\infty$, equation (\ref{Eq:50}) becomes:

\begin{equation}
\underbrace{\frac{\zeta c}{4\pi}
\frac{q_0^2}{2r_0}+\frac{\zeta
c}{4\pi}\frac{q_0^2}{2c\tau}}_{{\scriptsize{-\Delta mc^2}}}
=\underbrace{\frac{\zeta c}{4\pi}
\frac{q_0^2}{2r_0}}_{{\scriptsize{W_{Coul}
}}}+\underbrace{\frac{\zeta c}{4\pi}
\frac{q_0^2}{2c\tau}}_{{\scriptsize{W_{rad} }}}.
\label{Eq:51} 
\end{equation}

When a charge disappears according to Eq. (\ref{Eq:36}), its coulomb
field disappears as well. The energy of this field between the
surfaces of $r_0$ and $R$ decreases by the amount $\frac{\zeta c}{4\pi}
\left(\frac{q_0^2}{2r_0}-\frac{q_0^2}{2R}\right)$. As a result of the 
$\epsilon$-field (\ref{Eq:37}) action on the shell charges, their rest
energy increases by the amount: 

\[\Delta mc^2=-c\int_0^{\infty}q(t)\epsilon_{
\scriptscriptstyle\downarrow}(r_0,t)dt = \frac{\zeta c}{4\pi}
\frac{q_0^2}{2r_0}\frac{c\tau}{2r_0}\left(1-e^{-\frac{2r_{\mathrm
o}}{c\tau}}\right).\] 

Moreover, the energy given in (\ref{Eq:46}) radiates through the 
 surface $R$. The energy conservation law (\ref{Eq:48}), integrated over time
from 0 to $\infty$ is

\begin{equation}
\underbrace{\frac{\zeta c}{4\pi}\left(\frac{q_0^2}{2r_0} - 
\frac{q_0^2}{2R}\right)}_{{\scriptsize{-W_{Coul}}}}=
\nonumber
\end{equation}

%2345678901234567890123456789012345678901234567890123456789012345678901234567890
\begin{eqnarray}
=\underbrace{\frac{\zeta c}{4\pi}\frac{q_0^2}{2r_0}\frac{c\tau}{2r_0}
\left(1-e^{-\frac{2r_o}{c\tau}}\right)}_{{\scriptsize{\Delta mc^2}}} + 
\nonumber \\
\underbrace{\frac{\zeta c}{4\pi}\frac{q_0^2}{2r_0}
\left[1-\frac{c\tau}{2r_0} \left(1-e^{-\frac{2r_o}{c\tau}}\right)\right] - 
\frac{\zeta c}{4\pi}\frac{q_0^2}{2R}}_{{\scriptsize{W_{rad} }}}.
\label{Eq:52} 
\end{eqnarray}

We can see that at charge disappearance its coulomb field
energy transforms partly into rest energy and partly is emitted 
irretrievably. The de\-crea\-se of the coulomb energy exactly equal 
the sum of rest energy increase and the emitted energy.

In the case when $r_0\ll c\tau$ and $R\to\infty$, equation (\ref{Eq:52}) is
simplified:
\begin{equation}\underbrace{\frac{\zeta
c}{4\pi}\frac{q_0^2}{2r_0}}_{{\scriptsize{-W_{Coul} }}}
=\underbrace{\frac{\zeta c}{4\pi}\frac{q_0^2}{2r_0}- \frac{\zeta
c}{4\pi}\frac{q_0^2}{2c\tau}}_{{\scriptsize{\Delta
mc^2}}}+\underbrace{\frac{\zeta
c}{4\pi}\frac{q_0^2}{2c\tau}}_{{\scriptsize{W_{rad} }}}.
\label{Eq:53} 
\end{equation}

The above transitions of energy from one form into another look like 
the transitions of energy in an electric circuit: the electric
energy of a capacity transforms to the magnetic energy of inductance
and vice versa. Part of the energy is emitted in each such transition.
In our case (at the charge varying) rest energy passes to the 
coulomb field energy and vice versa. Part of the energy is emitted.

%2345678901234567890123456789012345678901234567890123456789012345678901234567890
\subsection{Wigner's paradox}
It was noted by Wigner that invariance considerations,
which substantiate the electric charge conservation law, are less
convincing than those which substantiate the conservation laws of
energy, momentum and angular momentum \cite{wigner}. In this
connection he tried to link conservation of charge with
conservation of energy. For this purpose Wigner considered the following 
thought experiment. Let us consider the charged Faraday cage 
with potential $\varphi_1$ and create a charge $q$ inside
it. Let the sign of the electric charge be the same as the sign of
the cage charge. Some amount of energy $W$ will be used for this. Then
we carry the charge into a point with potential $\varphi_2$, which is 
distant from the cage. The work which is done is  
$A=q(\varphi_1-\varphi_2)$. Now let us annihilate the charge. Thus we
will return the same energy $W$, as in Maxwell electrodynamics
none of the process depends on the actual 
potential value. Then we carry the  
particle, which is already uncharged, into a primary point inside the cage,
without spending energy for this movement.  During 
this closed cycle procedure we get work $A$, that contradicts 
the first law of thermodynamics.  From this consideration Wigner 
concludes that the initial assumption about the possibility of charge 
creation is wrong. But this consideration contains an obvious
logical error. 
Wigner's analysis of a process with a non-conserved charge
is based on Maxwell electrodynamics, which describes processes with
conserved charges only.  It is not necessary to discuss thought 
experiments with non-conserved charges in the framework of Maxwell equations 
in order to get a contradiction with fundamental physical 
statements. The equation $\nabla\cdot\bm{E}=\zeta c\rho$ shows that at 
a quite distant point from the charge the electric field appears or disappears
simultaneously with the appearance or disappearance of the charge.
So, it propagates with infinite velocity.

But there is no contradiction if processes with the non-conserved
charges are analysed with the help of equations (\ref{Eq:28})-(\ref{Eq:31}).

We can prove this by considering Wigner's circular process in the 
framework of Eq. (\ref{Eq:28})-(\ref{Eq:31}).
 While the charge is being created its rest energy decreases by the amount $\Delta
mc^2=W_{Coul}+W_{rad,1}$. This energy is used in order create 
the energy of the coulomb field $W_{Coul}$ and field of radiation
$W_{rad,1}$. Besides this, when the $\epsilon$-field, which appeared at the
moment of charge birth reaches the surface of Faraday cage, its
rest energy decreases by the amount $\Delta Mc^2=q\varphi_1$. This
follows from the fact that the cage section, having a charge 
$dQ$ and being at a distance of $r_{q,dQ}$ from the charge $q$,
decreases its rest energy by the amount 
$dQ\cdot (\zeta c)/(4\pi)\cdot q/(r_{q,dQ})$ according to (\ref{Eq:45}). 
The entire cage decreases its rest energy by the  amount

\[q\frac{\zeta c}{4\pi}\int_0^Q\frac{dQ}{r_{q,dQ}}=q\varphi_1.\]

Here $\varphi_1$ is the potential created by the cage at location of charge $q$.  
While transferring the charge to
a point with potential $\varphi_2$ we get the work
$A=q(\varphi_1-\varphi_2)$. At annihilation of the charge a part
of the energy of its Coulomb field $W_{Coul}$ is transformed into
radiation energy of $W_{rad,2}$, and the other part
$W_{Coul}-W_{rad,2}$ is transformed into rest energy and thus 
$mc^2-W_{Coul}-W_{rad,1}+(W_{Coul}-W_{rad,2})=mc^2-W_{rad,1}-W_{rad,2}$.
When the $\epsilon$-field, which appears at the charge annihilation,  
reaches the surface of cage, it initiates the increase of its
rest energy by $q\varphi_2$ and makes it equal to
$Mc^2-q\varphi_1+q\varphi_2$. Now we can return the uncharged
particle into its primary position without spending energy. As a
result of completion of the cycle, rest energy of a particle
decreases by $W_{rad,1}+W_{rad,2}$, and the rest energy of the cage
decreases by $q(\varphi_1-\varphi_2)$. The first decrease  
($W_{rad,1}+W_{rad,2}$) is spent for radiation of two
$\epsilon$-impulses at the birth and annihilation of the charge. The
second ($q\varphi_1-q\varphi_2$) --- for the work of moving the charge 
between potentials $\varphi_1$ and $\varphi_2$.  As we can see, 
the supposition about birth and annihilation of a charge,
discussed by equations (\ref{Eq:28})-(\ref{Eq:31}), does not 
contradict the energy conservation law.

%2345678901234567890123456789012345678901234567890123456789012345678901234567890
\subsection{Electromagnetic mass and renormalization}

As was shown in the previous section, a non-conserved electric charge does
not result in the non-conserving of energy. As the charge varies its
mass changes in a way that total energy (rest energy, coulomb field
energy and energy of radiation) remains constant.  Hence, at charge birth, 
rest energy is the source of field energy. It is not obvious 
that energy conservation must be provided exactly in this way.  
 It is possible to expect that energy of the
fields is created at the expense of energy of the external agent 
creating the charge. This is the exact Wigner's supposition 
\cite{wigner}.  However, equations (\ref{Eq:28})-(\ref{Eq:31}) are arranged in such a
way that they don't require any external agent.
According to  Eq. (\ref{Eq:50}) the non-field mass $m_e$ of the born charge
is less than the primary mass $m_0$ of the uncharged particle:

\begin{equation}
m_e=m_0-\frac{W_{Coul}}{c^2}-\frac{W_{rad}}{c^2}.
\label{Eq:54} 
\end{equation}

The non-field mass $m_e$ is the remainder of the primary mass $m_0$.
A part of the primary mass transforms into electromagnetic mass, which is 
equivalent to the coulomb field energy, and into radiation energy.  
 Total charge mass (non-field and electromagnetic) is
equal to:

\begin{equation}
m_e+\frac{W_{Coul}}{c^2}=m_0-\frac{W_{rad}}{c^2}.
\label{Eq:55} 
\end{equation}

When the radius of the charge tends to zero, $W_{Coul}$ and $m_e$ tend to
infinity but their sum (total mass) remains finite and less
than the primary mass of the un\-charg\-ed particle, because a part of
this mass transforms into energy of radiation.

Relationship (\ref{Eq:55}) is similar to that used for renormalization
of mass. The renormalization hypothesis assumes that observable
mass of the charged particle $m$ is the sum of the \lq\lq
naked\rq\rq\ unobservable mass $m_e$ and electromagnetic mass, and
this sum is a finite quantity:

\[m=m_e+\frac{W_{Coul}}{c^2}.\]

As we can see from the model described by equations (\ref{Eq:28})-(\ref{Eq:31}), such a 
relationship arises quite natural and consistently, but not as a 
supposition. The re\-normali\-za\-tion mechanism is not brought
from the outside, it is inherent to equa\-ti\-ons (\ref{Eq:28})-(\ref{Eq:31}), which
describe the processes of charge varying. Relationship
(\ref{Eq:55}) can be treated as substantiation  of the hypothesis of renormalization in
classical electrodynamics. It links the finite observable mass of
the charge with the primary mass of the uncharged particle. At the
same time, relationship (\ref{Eq:44}) may be treated as a dynamic  
realization of the renormalization hypothesis. It describes the
process of mass transference from the non-field form into the field
form and vice versa, at charge birth and annihilation.

From the above consideration
 it follows quite certainly, that mass $m$ from 
equations of motion (\ref{Eq:41}),(\ref{Eq:42}), decreases at charge birth and
increases at its annihilation. Therefore, it is the non-field mass.
Thus, these equations do not take into account the contribution of
electromagnetic mass into the charge inertia. However, there is no
doubt that the electromagnetic energy of the charge contributes to its
inertia. So the question arises: in what way must this contribution be taken
into account in the equations of motion? We also remember 
that equations (\ref{Eq:41}),(\ref{Eq:42}) do not take into account the influence
of their own field of radiation (radiation reaction) on the charge motion.  
Probably, both effects must be taken into account by the same 
mechanism: by introducing, into the equations of motion, the self-reaction
force of the $\bm{E}$ and $\bm{B}$ fields onto the
charge creating these fields.  Particle mass $m$ from the equations of 
motion (\ref{Eq:41}),(\ref{Eq:42})
decreases when its charge appears.   Also, self-reaction 
forces  must appear in these
equations. According to the Lorentz supposition \cite{l}, self-action
of a coulomb field must change the inertial properties of the charge and
in a such way compensate for the decreasing of the non-field mass. Therefore,
the contribution of electromagnetic mass must be taken into
account by means of the self-action of the coulomb field.  The self-action
of the transversal field must take into account the radiation reaction. This 
is an old problem of electrodynamics, which is not
solved yet.   Repeated attempts (for example, \cite{d,w1,w2}) to take 
self-action into account have failed.  
Our consideration allowed us to take into account the self-action of the 
$\epsilon$-field and to find out that this self-action results in
mutual transformations of the field and non-field masses at charge
varying.  However, the problem of the self-action of the $\bm{E}$ and
$\bm{B}$ fields remains unsolved.  A different method of accounting for 
the self-action of the coulomb field is proposed by renormalization.  
It is introducing, into the equations of motion, the complete
observable mass instead of the non-field mass and self-action of
coulomb field. It should be emphasized that renormalization is in no
way connected with the infinity of electromagnetic mass for a point
charge. Even if a method to make this mass finite is found,
the problem of including it into the equations of motion remains.

\section{Discussion}
The natural question is if it is really necessary to investigate processes
with non-conserved electric charge, when it is indeed conserved
in all known processes? There are as least two reasons for doing this.

The first reason is that such research helps us to understand better even
those questions which are not directly connected to charge
varying. Thus, for example, we find that there is a mechanism of
mutual transformations of rest energy and electric energy, which allows
a new look at the problem of renormalization and
self-action. For this reason it is important to extend
the research of non-conserved charge processes into the scope
of quantum electrodynamics.  It is interesting because  
in QED the Lorentz condition is executed only on 
average. As a consequence the $\epsilon$-field is also eliminated only
on average, $\langle\psi|\partial_\mu
A^\mu|\psi\rangle=\langle\psi|\epsilon|\psi\rangle=0$.  But it does
not eliminate the possibility of occurrence of the $\epsilon-\bm{E}$-field 
fluctuation effects.

The second reason is purely speculative. One can suppose the
violation of the electric charge conservation law in the area
of energies higher than attained now, and also in the early
stages of universe development. If processes with
charge varying are found at high energies  
 they will become the effective
method of rest energy transformation into electric energy.  Also, it is not
necessary to assume that elementary particles initially appeared
in their present form at the early stages of the  development of the universe. 
It is possible that originally matter was
maximally homogeneous and particles had only one property  ---
their masses. Subsequently differentiation took place, and particles
gain charges, spending part of their rest energy on the energy of field  
creation.

\section{Summary}
In this article we make the first step in research of the 
hypercomplex Dirac equation.  Already this first research
demonstrates interesting and unusual pro\-per\-ties of this
equation. In particular, this equation contains the
electrodynamics of non-conserved charges and describes the 
inter\-ac\-tion of such charges. These inter\-ac\-tion are highly  
unusual: they cause changes of rest mass and can be responsible for the
mechanism of renormalization of electric charge masses. However, 
many more questions remained unsolved. 

\subsection*{Acknowledgments}
The authors would like to thank Prof. Lukyanets S.P., Prof. Lev B.I., Prof. Tomchuk P.M. and Prof. Cooney for stimulating
discussions.

%2345678901234567890123456789012345678901234567890123456789012345678901234567890
\appendix*
\section{Conservation laws}
Combining the equations (\ref{Eq:08})-(\ref{Eq:15}), it is possible to get the
conservation laws of energy and momentum as in the Maxwell
electrodynamics. To get the conservation law of energy, let us
form a combination from equations (\ref{Eq:08})-(\ref{Eq:15})

\begin{eqnarray}
\epsilon\times (8)\! + \!{\bm E}\times (9)\! &-& \!\beta\times (10)\! + \nonumber \\ 
\!c{\bm B}\times (11)\! + \!V^0\times (12)\! &+& \!{\bm V}\times (13)\! + \nonumber \\ 
\!{\bm U}\times (14)\! &+& \!U^0\times (15). \nonumber  
\end{eqnarray}

After dividing by $\zeta$ we find

%\begin{gather}
\begin{align}
\frac{1}{c}\frac{\partial}{\partial
t}\frac{E^2+ c^2B^2+\epsilon^2+ \beta^2}{2\zeta}+\nonumber\\
\left({\bm E}\cdot{\bm j}_e- \epsilon c\rho_e -c{\bm B}\cdot{\bm j}_m-\beta
c\rho_m\right)+\nonumber\\
+\frac{1}{c}\frac{\partial}{\partial
t}\frac{V^2+U^2+(V^0)^2+(U^0)^2}{2\zeta}+ \nonumber\\
\left(\bm{V}\cdot\bm{k}-\bm{U}\cdot\bm{l}-
V^0s+U^0p\right)=\nonumber\\
=-\mathrm{div}\frac{\bm{E}\times c\bm{B}+\epsilon\bm{E}-\beta
c\bm{B}}{\zeta}- \nonumber\\
\mathrm{div}\frac{V^0\bm{V}+U^0\bm{U}+
\bm{V}\times\bm{U}}{\zeta}.
\label{Eq:56}
\end{align}
%\end{gather}

The sense of this relationship is obvious. Terms under 
$\partial/\partial t$ are the densities of energy of the fields,
terms with charges and currents are the rates of change of
energy of particles interacting with the fields. In other words,
it is the power of forces with which the fields affect charges and
currents. Terms under the sign of divergence are the
densities of energy flux being carried by the fields.

To get the conservation law of momentum, let us form the following 
combination from equations (\ref{Eq:08})-(\ref{Eq:15})

\begin{eqnarray}
 {\bm E}\times (8) + c{\bm B}\times (9) + c{\bm B}\times (10) &-& {\bm E}\times (11) + \nonumber \\ 
\epsilon\times (9)-\beta\times (11) + {\bm V}\times (14) &-& {\bm U}\times (13) - \nonumber \\ 
{\bm V}\times (12) - V^0\times (13) - U^0\times (14) &-& {\bm U}\times (15). \nonumber 
\end{eqnarray}

Dividing by $c\zeta$ yields:

\begin{eqnarray}
\frac{1}{c}\frac{\partial}{\partial t}
\frac{\bm{E}\times c\bm{B} - \epsilon\bm{E} + \beta c\bm{B}}{c\zeta} + \nonumber\\
+\frac{1}{c} \left(c\rho_e\bm{E}+\bm{j}_e\times c\bm{B} - c\rho_mc\bm{B} + \right. \nonumber\\
\left. \bm{j}_m\times \bm{E}-\epsilon\bm{j}_e-\beta\bm{j}_m\right) + \nonumber\\
+\frac{1}{c}\frac{\partial}{\partial t}\frac{\bm{U}\times\bm{V} + V^0\bm{V}+U^0\bm{U}}{c\zeta} + \nonumber\\
+\frac{1}{c}\left(-s\bm{V}+p\bm{U}+V^0\bm{k}- U^0\bm{l} +  \right. \nonumber\\
\left. \bm{U}\times\bm{k} + \bm{V}\times\bm{l}\right) = 
\nabla\cdot\mathsf{T} \nonumber\\
\label{Eq:57}
\end{eqnarray}

Here $\nabla\cdot\mathsf{T}$ is a divergence of the stress tensor

\begin{eqnarray}
\nabla\cdot\mathsf{T} = \nonumber \\
\frac{1}{c\zeta}\left[{\bm E}\nabla\cdot{\bm E} - {\bm E}\times\nabla\times{\bm E} + c{\bm B}\nabla\cdot c{\bm B} -  \right. \nonumber \\
\shoveright{\left. c{\bm B}\times\nabla\times c{\bm B} -\nabla\times(\epsilon c\bm{B}) - \right. \nonumber \\ 
\left. \nabla\times(\beta\bm{E}) + \epsilon\nabla\epsilon + \beta\nabla\beta\right]+}  \nonumber\\
+\frac{1}{c\zeta}\left[-{\bm V}\nabla\cdot{\bm V} + {\bm V}\times\nabla\times{\bm V} - {\bm U}\nabla\cdot{\bm U} +  \right. \nonumber \\
\left. {\bm U}\times\nabla\times{\bm U} + \nabla\times(V^0\bm{U}) - \nabla\times(U^0\bm{V}) - \right. \nonumber \\
\left. V^0\nabla V^0 - U^0\nabla U^0  \right].
\label{Eq:58}
\end{eqnarray}

The stress tensor is defined as

\begin{eqnarray}
T^{ij} =  \nonumber \\
\frac{1}{\zeta c}\left[E^iE^j\! + \!c^2B^iB^j\! - \!\frac{\eta^{ij}}{2}(E^2\! + \!c^2B^2\!-\!\epsilon^2\!-\!\beta^2)\! - \! \right. \nonumber \\
\left. \varepsilon^{ijk}(\beta E_k\!+\!\epsilon cB_k)\right]\! \nonumber \\
 -\frac{1}{\zeta c}\left[V^iV^j\! + \!U^iU^j\!+ \!\varepsilon^{ijk}(U^0V_k\!-\!V^0U_k)  - \!   \right. \nonumber \\
\left. \frac{\eta^{ij}}{2}\left(V^2\!+\!U^2\!-\!(V^0)^2\!-\!(U^0)^2 \right)\! \right]. \nonumber \\
\label{Eq:59}
\end{eqnarray}

In (\ref{Eq:57}) terms under $\partial/\partial t$ are the densities 
of the momentum of the fields, terms with charges and currents are the
densities of forces, with which the fields affect charges and
currents.   The stress tensor determines the density of the momentum flux
carried by the fields.

Our attention should be paid to the fact that the 
energy-momentum-stress tensor contains antisymmetric terms.
However, it can always be made symmetric by building an equivalent
metrical energy-momentum-stress tensor. 

%\nocite*
%\bibliography{artaREVTeX}
%\bibliographystyle{apsrev}

\end{document}